# Retrieval of nitric oxide in the mesosphere from SCIAMACHY nominal limb spectra

Stefan Bender[1], Miriam Sinnhuber[1], Martin Langowski[2], and John P. Burrows[3]

[1]Institute for Meteorology and Climate Research, Karlsruhe Institute of Technology, Karlsruhe, Germany
[2]Institute of Physics, Ernst–Moritz–Arndt–University of Greifswald, Greifswald, Germany
[3]Institute of Environmental Physics, University of Bremen, Bremen, Germany

*Correspondence to:* Stefan Bender (stefan.bender@kit.edu)

**Abstract.** We present a retrieval algorithm for nitric oxide (NO) number densities from measurements from the SCanning Imaging Absorption spectroMeter for Atmospheric CHartographY (SCIAMACHY, on Envisat) nominal limb mode (0–91 km). The NO number densities are derived from atmospheric emissions in the gamma bands in the range 230–300 nm, measured by the SCIAMACHY ultra-violet (UV) channel 1. The retrieval is adapted from the mesosphere and lower thermosphere mode (MLT, 50–150 km) NO retrieval (Bender et al., 2013), including the same 3-D ray tracing, 2-D retrieval grid, and regularisations with respect to altitude and latitude.

Since the nominal mode limb scans extend only to about 91 km, we use NO densities in the lower thermosphere (above 92 km), derived from empirical models, as a priori input. The priors are the Nitric Oxide Empirical Model (NOEM; Marsh et al., 2004) and a regression model derived from the MLT NO data comparison (Bender et al., 2015). Our algorithm yields plausible NO number densities from 60 to 85 km from the SCIAMACHY nominal limb mode scans. Using a priori input substantially reduces the incorrect attribution of NO from the lower thermosphere, where no direct limb measurements are available. The vertical resolution lies between 5 and 10 km in the altitude range 65–80 km.

Analysing all SCIAMACHY nominal limb scans provides almost 10 years (from August 2002 to April 2012) of daily NO measurements in this altitude range. This provides a unique data record of NO in the upper atmosphere and is invaluable for constraining NO in the mesosphere, in particular for testing and validating chemistry climate models during this period.

## 1 Introduction

Solar, auroral, and radiation belt electrons as well as soft solar X-rays produce nitric oxide (NO) in the mesosphere and lower thermosphere (MLT, 50–150 km). Hence, the NO content in this atmospheric region reveals how solar and geomagnetic activity and variability impact the atmospheric composition (Hendrickx et al., 2015; Sinnhuber et al., 2016). NO downward transport during polar winters then influences the lower atmosphere, in particular by catalytically reducing stratospheric ozone (Sinnhuber et al., 2012; Seppälä et al., 2015; Verronen and Lehmann, 2015). This, in turn, changes the heating and cooling and eventually the atmospheric circulation, possibly down to tropospheric weather systems as supported by observations (Seppälä et al., 2009; Maliniemi et al., 2014). Therefore, NO data from the mesosphere are crucial for linking solar and geomagnetic activity to atmospheric composition and dynamics.

We adapt the NO retrieval for the SCIAMACHY (SCanning Imaging Absorption spectroMeter for Atmospheric CHartographY) MLT limb mode (Bender et al., 2013) to the nominal limb scans. Since the nominal scans are carried out only up to about 91 km, they do not sample the lower thermosphere where the largest NO densities are typically located, above ≈ 100 km (Barth et al., 2003; Marsh et al., 2004; Minschwaner et al., 2004; Funke et al., 2005; Bender et al., 2015). The instrument observes all emissions along the line of sight, including the NO layer in the lower thermosphere. However, because the nominal limb scans have no tangent points above 91 km, the retrieval maps these enhanced values to lower altitudes. We counter these distorted values by using non-zero a priori input in the lower thermosphere above 91 km.

Exactly as in the case of the MLT retrieval, we use SCIAMACHY's UV (ultra-violet) channel 1 (214–334 nm) to derive the NO number densities from atmospheric emissions of the NO gamma bands. Retrieving NO from the gamma band emissions was done previously in a variety of rocket (Cleary, 1986; Eparvier and Barth, 1992) and satellite (Frederick and Serafino, 1985; Barth et al., 2003; Minschwaner et al., 2004) experiments.

Further NO measurements in the mesosphere using different emissions were carried out, for example, using the Sub-Millimetre Radiometer (SMR; Pérot et al., 2014) and the The Optical Spectrograph and InfraRed Imaging Spectrometer (OSIRIS; Sheese et al., 2011, 2013) on board the Odin satellite. Infrared limb emission measurements were also performed, using the Michelson Interferometer for Passive Atmospheric Sounding (MIPAS) on the Envisat satellite (Funke et al., 2005; Bermejo-Pantaleón et al., 2011). The Atmospheric Chemistry Experiment Fourier transform spectrometer (ACE-FTS) on board SCISAT-1 observed NO via solar occultations at infrared wavelengths (validated in Kerzenmacher et al., 2008). More information about previous measurements of NO in the mesosphere can be found in Bender et al. (2013).

This paper is organised as follows: we present some details about SCIAMACHY and its nominal limb mode in Sect. 2. Section 3 goes into detail about the retrieval method. Finally, we present our results and discuss the influence of different a priori choices in Sect. 4. Conclusions are given in Sect. 5.

## 2 SCIAMACHY nominal mode

SCIAMACHY is a UV–visible–near-infrared (214–2380 nm) spectrometer on the European Envisat satellite. This satellite was in a sun-synchronous orbit at approximately 800 km altitude from 2002 (Burrows et al., 1995; Bovensmann et al., 1999) but communication was lost in April 2012. The measurement modes include limb and nadir sounding as well as lunar and solar occultation measurements. Compared to the twice monthly driven MLT mode, described in Bender et al. (2013) and Langowski et al. (2014), the nominal limb scanning mode of SCIAMACHY was performed almost daily for almost 10 years of successful Envisat operations. The nominal mode comprises limb scans from about −4 up to 91 km, which is just short of the maximum NO number density in the lower thermosphere. However, since this mode was carried out almost daily, it delivered a unique data set of global continuous spectral measurements of the whole mesosphere.

From the beginning of its mission in August 2002 to the middle of October 2003, the nominal limb scans extended up to 105 km. On 15 October 2003, the sampling pattern of the nominal mode was changed to 30 limb tangent points from −4.5 to about 91 km in approximately 3.3 km steps. The horizontal distance between the individual limb scans is about 7–8°, amounting to about 24 usable limb scans on the dayside of the orbit. More details about the measurement orbit sequence of SCIAMACHY can be found in Bovensmann et al. (1999).

As in the case of the SCIAMACHY MLT retrieval (Bender et al., 2013), our retrieval is based on the NO gamma band emissions. These are fluorescent emissions of which the first electronic state is excited by solar UV radiation, thus restricting the useful measurements to daytime.

## 3 Retrieval algorithm

The retrieval is adapted from the SCIAMACHY MLT NO retrieval; see Bender et al. (2013); Scharringhausen et al. (2008a, b); Langowski et al. (2014). We use the limb tangent point spectra from 50 km to the top at about 91 km. From these we fit the measured NO gamma bands spectra to modelled spectra (Eparvier and Barth, 1992; Stevens, 1995) to determine the NO slant column densities along the line of sight.

As discussed in Eparvier and Barth (1992); Stevens (1995); Bender et al. (2013), the NO gamma bands in the UV are fluorescent emissions arising from an electronic transition from the first excited state $A^2\Sigma$ to the ground state $X^2\Pi$. Superposed on this electronic transition are the vibrational and rotational transitions, which constitute the different gamma bands.

The SCIAMACHY spectrometer resolves the vibrational lines but not the rotational lines. An example of a synthetic spectrum in the SCIAMACHY UV wavelength range can be found in Bender et al. (2013). As in the case of the MLT NO retrieval, we employ the (0, 2), (1, 4), and (1, 5) vibrational transitions to retrieve the NO number densities from the nominal mode limb spectra. As discussed in Bender et al. (2013), the line of sight from the emission point to the satellite instrument is optically thin in these cases. An optically thin line of sight substantially decreases the computational effort for the forward model. Furthermore, we calculate the emissivities of the NO gamma bands with the same parameters as given in Bender et al. (2013).

### 3.1 Radiative transfer

A generic objective in atmospheric remote sounding is to extract the relevant values (almost always at places without direct measurements) from the measured quantities. SCIAMACHY, and any limb sounder in general, measures the electromagnetic spectra, reaching the instrument from along the line of sight through the atmosphere. In this study, we aim to derive the NO number density distribution in the mesosphere.

The general forward model $F$ relates measurements $\boldsymbol{y}$ to their depending unknown quantities $\boldsymbol{x}$ via

$$F : \text{unknowns} \to \text{measurements}, \quad \boldsymbol{x} \mapsto \boldsymbol{y} = F(\boldsymbol{x}). \quad (1)$$

In general, the measured values are the spectral intensities at the satellite point. However, since we use non-resonant gamma transitions, these intensities are linearly related to the slant column densities (Bender et al., 2013). In our case, the measurements $\boldsymbol{y}$ are therefore given by the slant column densities by fitting the calculated NO gamma band spectra to the measured limb spectra. As noted above, $\boldsymbol{x}(s)$ are the number densities of NO at our self-defined retrieval grid point $s$.

Our forward model calculates the line of sight fraction within each grid cell and accounts for ozone and weak oxygen absorption along the line of sight from the grid point to the satellite and along the line from the sun to the grid point. Knowing the length of the line of sight within the grid cell, we can then calculate the possible emission signal from this cell. The details of this calculation are laid out in Scharringhausen et al. (2008b) and in Langowski et al. (2014).

Fitting each vibrational NO gamma transition individually yields three separate values for the slant column densities and their uncertainties at the tangent points. Thus, $\boldsymbol{y} \in \mathbb{R}^{3N_{\text{TP}}}$ (with $N_{\text{TP}}$ the number of tangent points) and is used in this way to retrieve the number densities $\boldsymbol{x}$.

The electronic excitation and emission occur at different wavelengths in general and for the three chosen gamma bands in particular. To account for this, we calculate the ozone and oxygen absorption-corrected emission matrix for the three gamma bands separately.

We subtract the Rayleigh background before the spectral fit. This enables NO number densities to be retrieved for cases in which the spectra are contaminated by polar mesospheric clouds (PMCs), also known as noctilucent clouds (NLCs).

### 3.2 Retrieval algorithm

In our case, $\boldsymbol{x}$ and $\boldsymbol{y}$ in Eq. (1) have different dimensions, in particular, $\dim(\boldsymbol{x}) > \dim(\boldsymbol{y})$. Therefore, Eq. (1) can not be solved exactly because it has infinitely many solutions (assuming the system is consistent).

For one orbit, the SCIAMACHY MLT limb scans involve approximately $\dim(\boldsymbol{y}) \approx 25 \times 30 \times 3 = 2250$ measurements (limb scans per orbit × tangent points per limb scan × separate spectral fits). Whereas one orbit of SCIAMACHY nominal limb scans delivers $\dim(\boldsymbol{y}) \approx 25 \times 13 \times 3 = 975$ measurements above 50 km. The retrieval is performed on a 2.5° × 2 km grid from 90° S to 90° N and from 60 to 160 km. The grid size is $\dim(\boldsymbol{x}) = 72 \times 51 = 3672$ (latitude × altitude), clearly larger than $\dim(\boldsymbol{y})$.

**Table 1.** Regularisation parameters as used in the NO retrieval from SCIAMACHY limb scans.

| $\lambda_{\text{a}}$ | $\lambda_{\text{alt}}$ | $\lambda_{\text{lat}}$ |
|---|---|---|
| $1 \times 10^{-17}$ | $3 \times 10^{-17}$ | $1 \times 10^{-16}$ |

Underdetermined systems can be solved using additional constraints, such as a priori input and regularisations (Rodgers, 1976; von Clarmann et al., 2003; Funke et al., 2005). Here we apply both prior input as well as vertical and horizontal regularisation. The a priori NO number densities are denoted by $\boldsymbol{x}_{\text{a}}$ and the accompanying covariance matrix by $\mathbf{S}_{\text{a}}$. We denote the regularisation matrices by $\mathbf{R}_{\text{alt}}$ and $\mathbf{R}_{\text{lat}}$; see Scharringhausen et al. (2008a); Langowski et al. (2014); Bender et al. (2013). We refer to Bender et al. (2013) for the exact details of how these are set up. The corresponding weighting parameters are denoted by $\lambda_{\text{a}}$, $\lambda_{\text{alt}}$, and $\lambda_{\text{lat}}$ for the a priori covariances, and the vertical and horizontal regularisation; they are listed in Table 1[1].

The retrieval is equal to minimising the regularised $\chi^2$:

$$\chi^2_{\text{reg}} = \| \boldsymbol{y} - \mathbf{K}\boldsymbol{x} \|^2_{\mathbf{S}_y^{-1}} + \| \boldsymbol{x} - \alpha\boldsymbol{x}_{\text{a}} \|^2_{\mathbf{S}_{\text{a}}^{-1}}$$
$$+ \lambda_{\text{alt}} \| \mathbf{R}_{\text{alt}}\boldsymbol{x} \|^2 + \lambda_{\text{lat}} \| \mathbf{R}_{\text{lat}}\boldsymbol{x} \|^2 , \quad (2)$$

where $\mathbf{K}$ is the Jacobian of Eq. (1) and $\alpha$ is an additional scale factor for the a priori densities. We minimise Eq. (2)[2] with the same iterative algorithm as described in von Clarmann et al. (2003); Funke et al. (2005). For $\alpha = 1$ (no a priori scaling) the intermediate solution is given by

$$\boldsymbol{x}_{i+1} = \boldsymbol{x}_i + \left( \mathbf{K}^T \mathbf{S}_y^{-1} \mathbf{K} + \mathbf{S}_{\text{a}}^{-1} + \mathbf{R} \right)^{-1}$$
$$\cdot \left[ \mathbf{K}^T \mathbf{S}_y^{-1} (\boldsymbol{y} - \boldsymbol{y}_i(\boldsymbol{x}_i)) \right.$$
$$\left. + \mathbf{S}_{\text{a}}^{-1} (\boldsymbol{x}_{\text{a}} - \boldsymbol{x}_i) - \mathbf{R}\boldsymbol{x}_i \right] , \quad (3)$$

with the regularisation matrix $\mathbf{R}$ defined as

$$\mathbf{R} := \lambda_{\text{alt}} \mathbf{R}_{\text{alt}}^T \mathbf{R}_{\text{alt}} + \lambda_{\text{lat}} \mathbf{R}_{\text{lat}}^T \mathbf{R}_{\text{lat}} . \quad (4)$$

Including the a priori scale $\alpha$ as a free parameter within a combined vector $(\boldsymbol{x}, \alpha)^T$, the retrieval step changes to

$$\begin{pmatrix} \boldsymbol{x} \\ \alpha \end{pmatrix}_{i+1} = \begin{pmatrix} \boldsymbol{x} \\ \alpha \end{pmatrix}_i + \begin{pmatrix} \mathbf{K}^T \mathbf{S}_y^{-1} \mathbf{K} + \mathbf{S}_{\text{a}}^{-1} + \mathbf{R} & -\mathbf{S}_{\text{a}}^{-1} \boldsymbol{x}_{\text{a}} \\ -\boldsymbol{x}_{\text{a}}^T \mathbf{S}_{\text{a}}^{-1} & \boldsymbol{x}_{\text{a}}^T \mathbf{S}_{\text{a}}^{-1} \boldsymbol{x}_{\text{a}} \end{pmatrix}^{-1}$$
$$\cdot \begin{pmatrix} \mathbf{K}^T \mathbf{S}_y^{-1} (\boldsymbol{y} - \boldsymbol{y}_i(\boldsymbol{x}_i)) + \mathbf{S}_{\text{a}}^{-1} (\alpha_i \boldsymbol{x}_{\text{a}} - \boldsymbol{x}_i) - \mathbf{R}\boldsymbol{x}_i \\ \boldsymbol{x}_{\text{a}}^T \mathbf{S}_{\text{a}}^{-1} (\boldsymbol{x}_i - \alpha_i \boldsymbol{x}_{\text{a}}) \end{pmatrix} . \quad (5)$$

---

[1]Note that, because of improved so-called M-factors (a measure of the degradation of the SCIAMACHY detectors) and an adjusted fit error calculation, the values reported here differ from the ones given in (Bender et al., 2013).

[2]In this notation the subscript matrix acts like a metric for the norm: $\| \boldsymbol{x} \|^2_M = \boldsymbol{x}^T \mathbf{M} \boldsymbol{x}$.

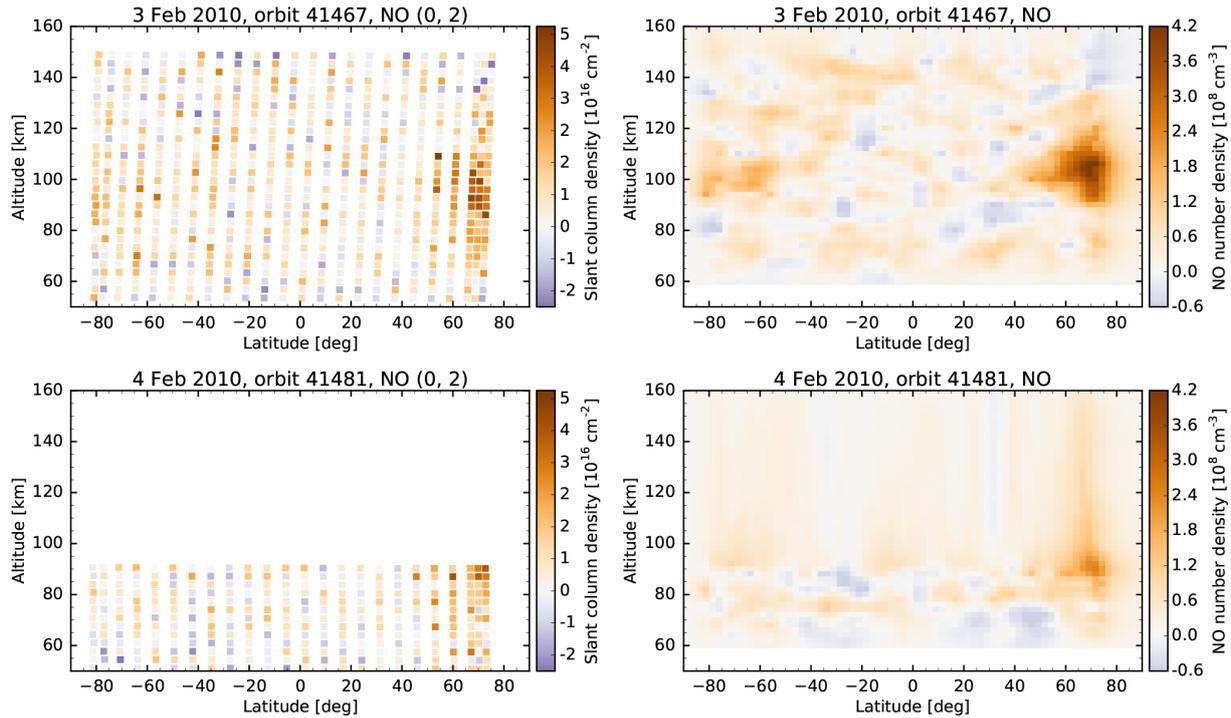

**Figure 1.** NO slant column densities from the (0, 2) gamma band (left column) and retrieved number densities from all used gamma bands (right column) along one SCIAMACHY MLT orbit (no. 41 467, 3 February 2010, top row) and one nominal orbit (no. 41 481, 4 February 2010, bottom row).

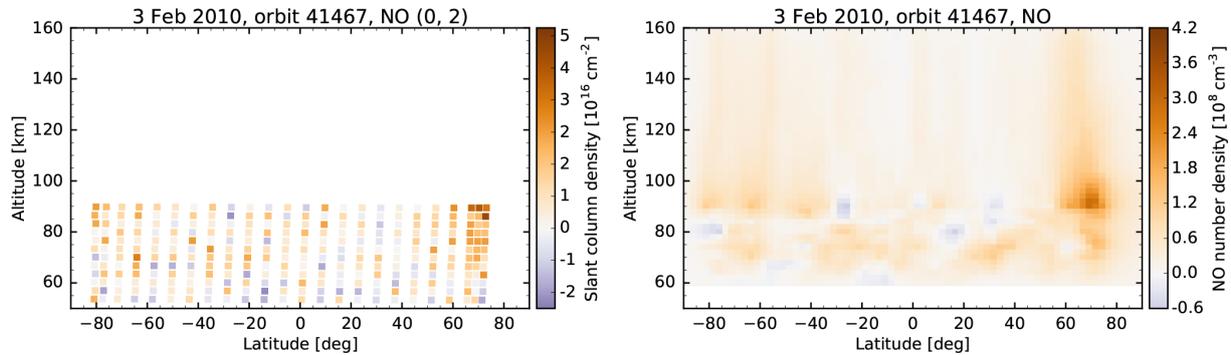

**Figure 2.** Results from the SCIAMACHY MLT sample orbit (no. 41 467, 3 February 2010) restricted to the 50–91 km limb tangent points. NO slant column densities for the (0, 2) transition (left) and the retrieved NO number densities without a priori input (right).

### 3.3 A priori density set-up

The SCIAMACHY nominal limb scans extend only up to ≈ 91 km in altitude, but the maximal NO density is typically located between 100 and 110 km in polar regions (Barth et al., 2003; Marsh et al., 2004; Funke et al., 2005; Bender et al., 2015). This maximal density lies along the line of sight of the instrument but it is not explicitly resolved with dedicated limb tangent points. Retrieving the NO number densities up to 160 km, these missing measurements adversely affect all values, in particular those below 100 km. Using a priori input to model these maximal values in the lower thermosphere is therefore necessary to retrieve correct number densities below 100 km.

In this work, we employ two models of the NO number density in the thermosphere. The first is the Nitric Oxide Empirical Model (NOEM; Marsh et al., 2004), an empirical model based on Student Nitric Oxide Explorer (SNOE) observations (Barth et al., 2003). This model is also used as a priori input for the MIPAS NO retrievals (Funke et al., 2005). It is, however, constructed only for altitudes from 100 to 150 km (Marsh et al., 2004) and set to constant below 100 km.

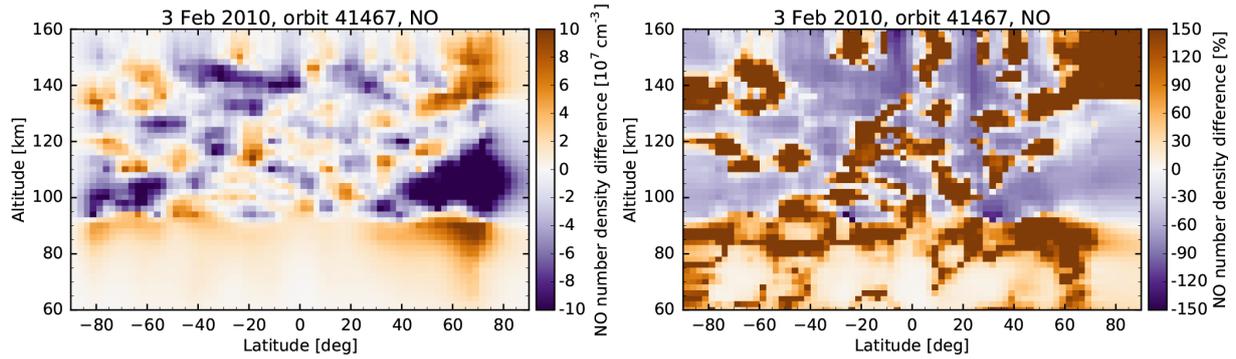

**Figure 3.** Absolute NO number density differences (left) and relative differences (right) from the SCIAMACHY MLT mode limb scans restricted to 50–91 km and zero a priori input compared to the full retrieval (sample orbit no. 41 467, 3 February 2010).

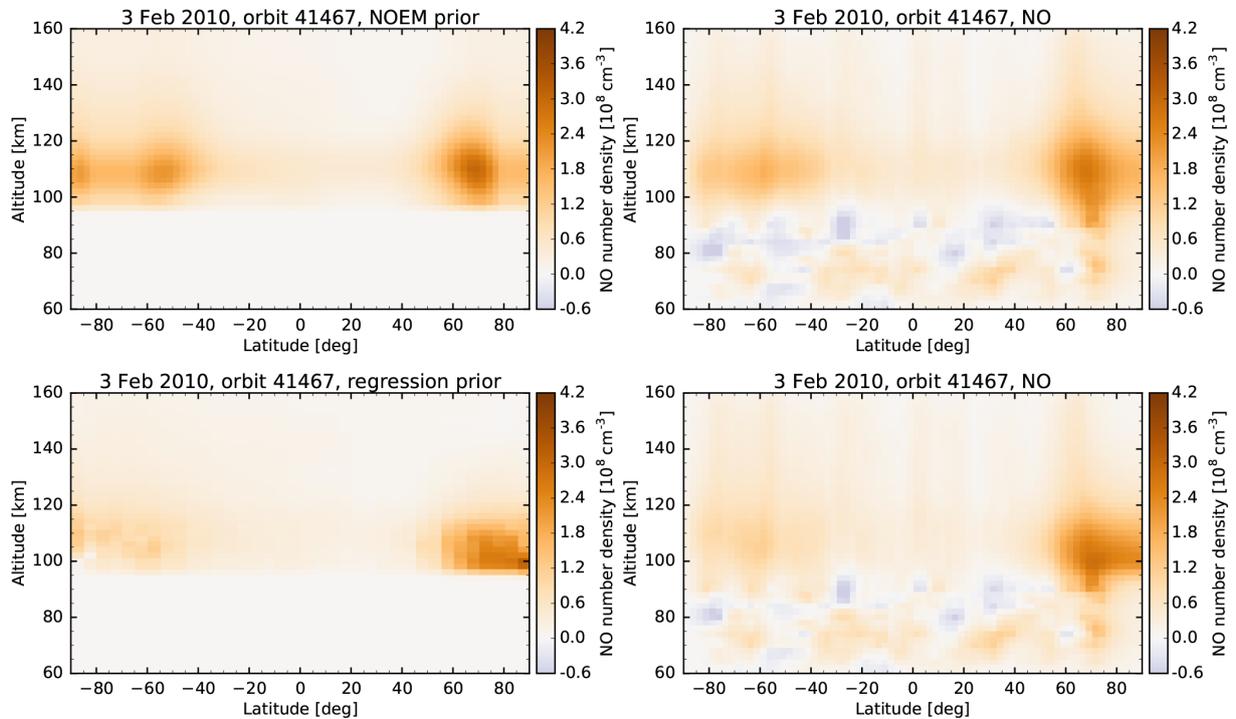

**Figure 4.** NO number density a priori input from the NOEM model (upper left) and the regression model (lower left). Retrieved NO number densities from the limited SCIAMACHY MLT limb scans using the NOEM model (upper right) and the regression model (lower right) a priori input (sample orbit no. 41 467, 3 February 2010).

The second model uses results from the multilinear regression analysis from the measurement comparison study of NO from four instruments: ACE-FTS, MIPAS, SCIAMACHY, and SMR (Bender et al., 2015). Here we use the parameters from the regression fit from 94 to 150 km of the composite data set.

At 92 km and below, the prior is set to zero with a smooth transition between 100 and 92 km. For the smooth transition we apply the standard analysis approach for the partition of unity. First, we define the smooth function $\theta(t)$ by

$$\theta(t) := \begin{cases} 0, & t \leq 0 \\ e^{-1/t}, & t > 0 \end{cases}, \quad (6)$$

and the smooth transition $\phi(s)$ from zero to one by

$$\phi(s) := \frac{\theta(s)}{\theta(s) + \theta(1-s)} = \begin{cases} 0, & s \leq 0 \\ 1, & s \geq 1 \end{cases}. \quad (7)$$

Note that $\phi(s)$ is smooth for all $s \in \mathbb{R}$. Then a smooth transition from one at altitude $a$ to zero at and below $a - w$ is given

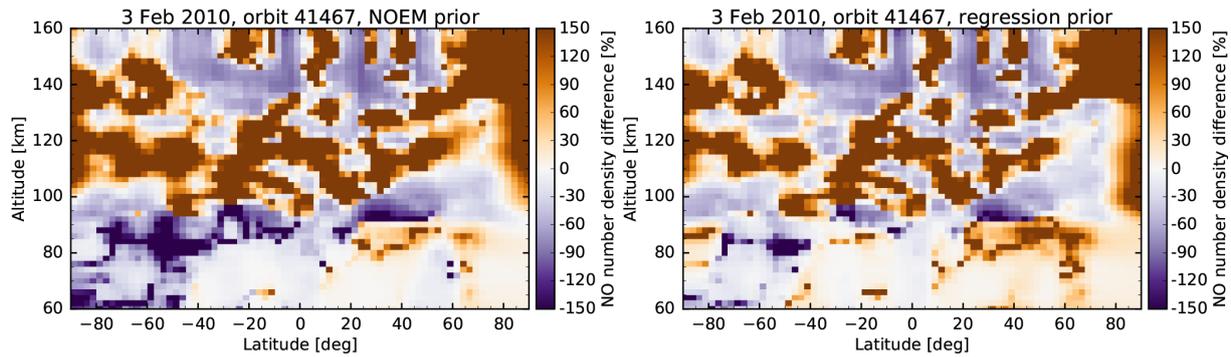

**Figure 5.** Relative NO number density differences from the SCIAMACHY MLT mode limb scans restricted to 50–91 km using the NOEM model (right) and the regression model (left) as a priori input compared to the full retrieval (sample orbit no. 41 467, 3 February 2010).

by scaling the argument of $\phi$ in the following way:

$$\Phi(s,a,w) := \phi\left(\frac{s-a+w}{w}\right) = \begin{cases} 0, & s \leq a-w \\ 1, & s \geq a \end{cases}. \quad (8)$$

We eventually scale the a priori input $x_a$ by $\Phi(s,a,w)$ with $a = 100\,\text{km}$ and $w = 8\,\text{km}$.

### 3.4 Further model input

As described in Bender et al. (2013), the atmospheric temperatures are needed to calculate the emissivity of the NO transitions. We use the NRLMSISE-00 model (Picone et al., 2002), which derives the temperatures using the geomagnetic $A_p$ index and the solar radio flux $f_{10.7}$. Both are taken as daily values from the Space Physics Interactive Data Resource (SPIDR) service (NGDC and NOAA, 2011) of the National Geophysical Data Center (NGDC) of the National Oceanic and Atmospheric Administration (NOAA).

## 4 Results

### 4.1 NO number density

As in the MLT retrieval described in Bender et al. (2013), we present the individual steps of the retrieval on the basis of the same sample orbit as above, number 41 467 from 3 February 2010. We chose this orbit because it shows the density maximum at high northern latitudes around 105 km. It is therefore a clear example of how the missing measurements above 91 km influence the results below. The upper-left panel of Fig. 1 shows the fitted NO slant column densities from the (0, 2) band at all tangent points of the example MLT orbit. The slant columns from the other bands look similar but have a worse signal to noise ratio; they are shown in Bender et al. (2013). As also discussed there, the region of largest slant column densities lies between 60 and 70° N around 100 km.

Accordingly, using the MLT retrieval algorithm without a priori input, the retrieved NO number densities for this example orbit are shown in the upper-right panel of Fig. 1. In addition to the expected high-density region in the Northern Hemisphere between 55 and 75° N and around 105 km, we also observe enhanced number densities in the Southern Hemisphere. There we find higher values between 55 and 85° S around 100 km.

In contrast, the lower-left panel of Fig. 1 shows the slant column densities of a sample orbit containing nominal limb scans (no. 41 481 from 4 February 2010). This orbit covers approximately the same longitude as the chosen MLT orbit above, but on the following day. As discussed above, the highest tangent point lies at about 91 km which manifests in shorter columns in the lower-left panel of Fig. 1 compared to the upperleft panel of the same figure.

The retrieved NO number densities from these slant column densities are shown in the lower-right panel of Fig. 1. Since we extended the retrieval up to 160 km such that the results are directly comparable to the MLT limb scan results, values above about 91 km smoothly extend upwards. This is a direct consequence of the applied regularisation. In general, the number densities are smaller than from the MLT orbit the day before. We assume that these lower values are related to the large natural variability of NO in the mesosphere.

### 4.2 Simulation of nominal mode

SCIAMACHY's MLT mode and nominal mode were carried out on different days. To estimate the error introduced by only sampling up to 91 km instead of 150 km, we simulate the nominal mode on MLT days by restricting the limb scan data from the MLT mode to 91 km before performing the retrieval. We then compare the resulting number densities over the full altitude range. However, when evaluating the results the main focus should be on altitudes below 91 km. This comparison begins with the results without a priori input and then evaluates the impact of the different prior choices below in Sect. 4.3.

The restricted slant column densities are shown in the left panel of Fig. 2. The retrieved NO number densities from these slant column densities are shown in the right panel of Fig. 2.

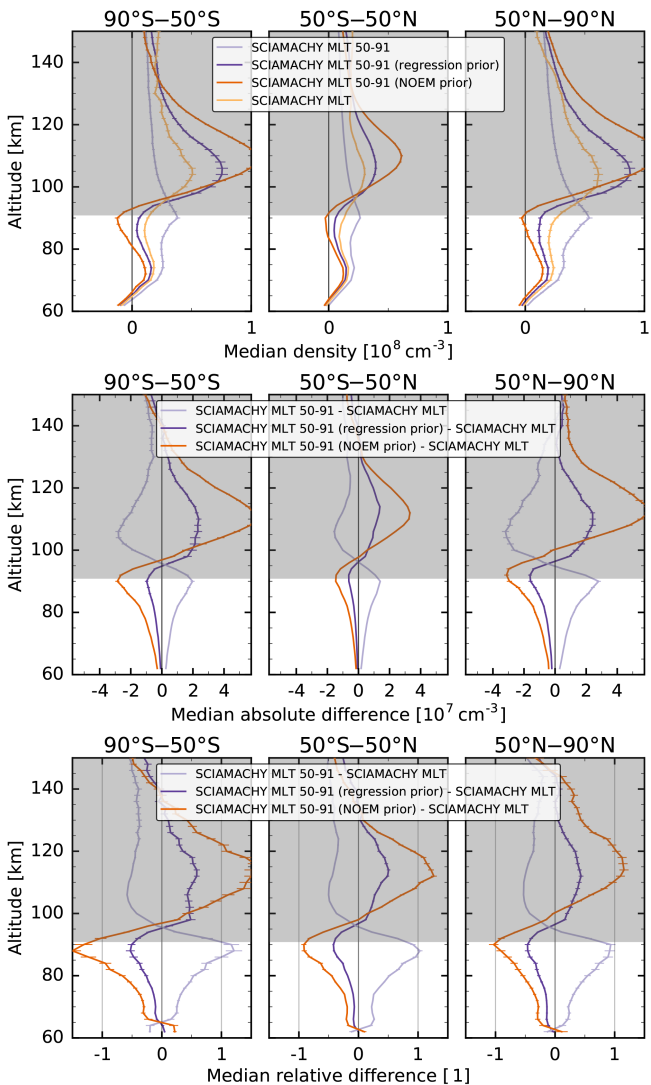

**Figure 6.** Median of the NO number densities (top) as well as the absolute (middle) and relative (bottom) differences to the full MLT retrieval in three latitude bands. The results are shown for the restricted (50–91 km) MLT scans without a priori input (light purple), and using the two a priori choices (dark purple is regression model, dark orange is the NOEM model). The number densities from the full SCIAMACHY MLT scans are shown in light orange.

Figure 3 shows the absolute (left panel) and relative (right panel) differences of the retrieved NO number densities without a priori input compared to the full MLT retrieval. The differences are mostly negative above 91 km because there we have no slant column densities available. Equivalently, the differences are positive below this altitude for the same reason: the retrieval algorithm incorrectly attributes the measured slant column densities to NO number densities at lower altitudes. The sometimes large relative differences should not be over-interpreted because at places with low number densities, large relative differences occur easily.

### 4.3 A priori influence

As seen in Fig. 3, omitting the limb tangent points above 91 km enhances the retrieved NO number densities below about 91 km. To compensate for these enhanced values, we used non-zero a priori input to model the NO number density above 100 km, as described in Sect. 3.3. The a priori input from the NOEM model (Marsh et al., 2004) is shown in the upper-left panel of Fig. 4 for the chosen example orbit. Similarly, the input from the regression model is shown in the lower-left panel of Fig. 4.

Including these prior inputs, the retrieved NO number densities from the restricted MLT limb scans are shown in the right column of Fig. 4. Therein, the upper-right panel shows the results using the NOEM model and the lower-right panel the results using the regression a priori densities.

Figure 5 shows the relative differences of the retrieved NO number densities from the restricted MLT limb scans compared to the full MLT retrieval results, this time using a priori input (compare to Fig. 3). In the first figure the NOEM model served as input above 100 km, and in the second case the input was derived from the regression analysis (Bender et al., 2015).

Figure 5 gives only a rough qualitative look at the differences when using the different a priori models. When compared to using no a priori input (Fig. 3), we find that below 90 km the NOEM a priori model (left panel of Fig. 5) results in much lower differences in the Northern Hemisphere, but apparently large negative differences in the Southern Hemisphere. Using the regression model as prior input (right panel of Fig. 5), a similar north–south gradient appears with slightly larger positive differences at northern latitudes and smaller negative differences in the Southern Hemisphere compared to the NOEM prior. In both cases the number densities are closer to the full MLT retrieval results than without a priori densities. Note, however, that the above comparison shows only a single orbit. To assess the differences quantitatively, we calculate the differences of the respective daily zonal mean number densities for all SCIAMACHY MLT measurement days. In particular, we compare the daily mean number densities in three latitude bands: 90–50° S, 50° S–50° N, and 50–90° N.

The zonal median number densities retrieved from the full MLT scans as well as the restricted MLT scans with and without a priori information above 90 km are shown in the top panel of Fig. 6. As above, restricting the covered altitude range to 91 km results in overestimating the number densities between 70 and 91 km. Using either a priori input underestimates the number densities in that altitude region more severely when using the NOEM model. In the latter case, this can even result in negative median number densities.

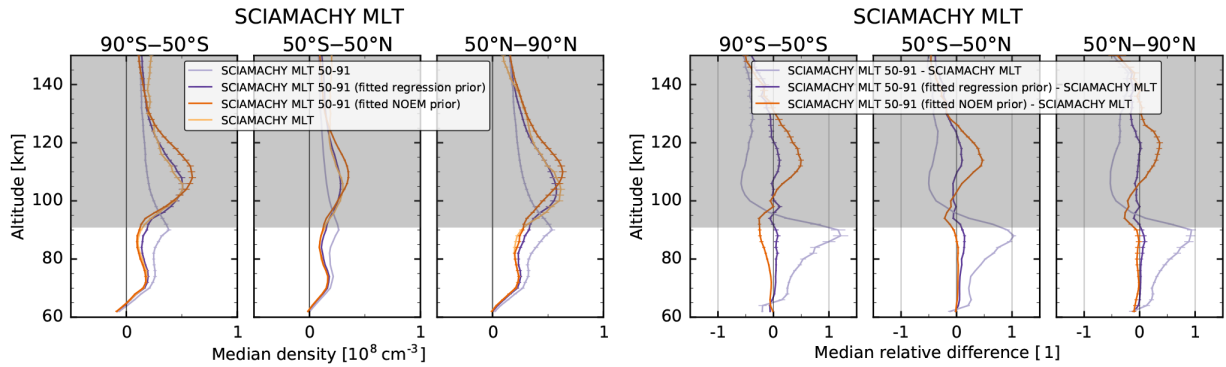

**Figure 7.** Median of the NO number densities (left) and the relative differences (right) using the scaled a priori values compared to the full SCIAMACHY MLT limb scan number densities.

**Table 2.** Differences of the NO number densities using the different a priori choice compared to the full MLT limb scan retrieved values. NH and SH indicate exceptions at high northern and high southern latitudes.

| Altitude range | Zero | NOEM | Regression | Scaled NOEM | Scaled regression |
|---|---|---|---|---|---|
| 85–91 km | 90…120 % | −70…−100 %, −90…−150 % (SH) | −40…−50 % | < ±10 %, −20…−25 % (SH) | 5…15 % |
| 75–85 km | 30…75 % | −40…−70 %, −45…−90 % (SH) | −20…−35 % | < ±5 %, −5…−20 % (SH) | 5…10 % |
| < 75 km | < 30 % | > −40 % | > −20 % | < ±3 %, 3…5 % (SH) | 3…7 % |

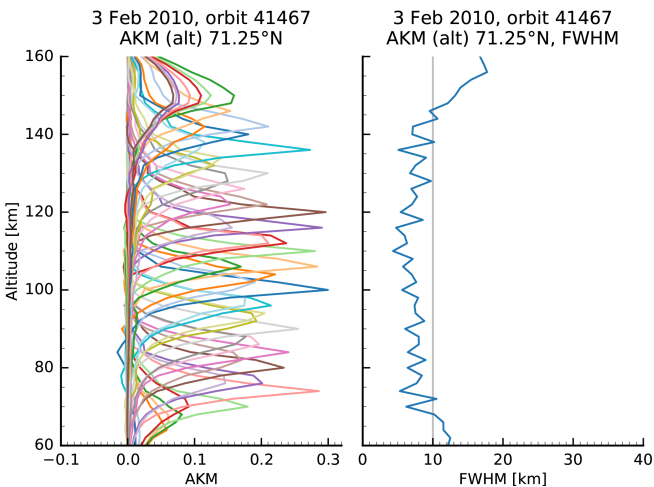

**Figure 8.** Altitude averaging kernel matrix elements (left panel) for a sample orbit (no. 41 467, 3 February 2010) at a particular latitude grid point (71.25° N). Corresponding full width at half maximum values (right panel) are shown for the same location.

The median of the absolute differences of the number densities of restricted MLT scans to the results from the full MLT retrieval are shown in the middle panel of Fig. 6. The bottom panel of Fig. 6 shows the median of the corresponding relative differences, illustrating the effect of the different a priori inputs.

In the following we focus on the differences between 60 and 91 km. Without a priori input (light blue line in Fig. 6), the retrieved number densities at the top most altitude (91 km) are up to $1 \times 10^7$ cm$^{-3}$ (middle and low latitudes) and up to $2…3 \times 10^7$ cm$^{-3}$ larger (high latitudes). In all three regions, this corresponds to about 100 %. Below about 80 km at high latitudes, the difference is smaller than $1 \times 10^7$ cm$^{-3}$, which translates to about 50 %. At middle and low latitudes, the absolute differences drop to about $0.5 \times 10^7$ cm$^{-3}$ at 80 km, which is equal to about 50 %. Below 70 km, the differences drop further, to below about 25 % in all regions.

Using a priori inputs (dark orange and dark purple lines in Fig. 6) yields smaller number densities below 91 km in general. Compared to the full MLT retrieval, using the NOEM model (dark orange line in those figures) has the largest effect with $3 \times 10^7$ cm$^{-3}$ ($\approx 100$ % (NH), $\approx 150$ % (SH)) smaller number densities at high latitudes at 91 km. Below 85 km in these regions, the difference drops to about $-2 \times 10^7$ cm$^{-3}$ ($\approx -70$ (NH), $\approx -100$ % (SH)) and further to $-1 \times 10^7$ cm$^{-3}$ ($\approx -40$ %) below 75 km. At middle and low latitudes, the largest difference is about $-2 \times 10^7$ cm$^{-3}$ ($\approx -100$ %) at 91 km. Below 85 km in this region, the difference declines to less than about $-1 \times 10^7$ cm$^{-3}$ ($\approx -75$ %), and below 75 km further to about $-0.5 \times 10^7$ cm$^{-3}$ ($\approx -20$ %)

Using the regression model as a priori input (dark purple line), these differences are smaller. At high latitudes, the largest difference is $-2 \times 10^7$ cm$^{-3}$ (−40 to −50 %) at 91 km, dropping to less than $-1 \times 10^7$ cm$^{-3}$ (−40 %) below about 85 km. Below 75 km at high latitudes, the differences become smaller than about $-0.5 \times 10^7$ cm$^{-3}$ (−20 %). At middle and low latitudes, the largest difference is approx-imately $-1 \times 10^7$ cm$^{-3}$ ($-50\%$) at 91 km, declining to less than $-0.3 \times 10^7$ cm$^{-3}$ ($-25\%$) below 85 km. Below 75 km in this region, the difference is smaller than $-0.15 \times 10^7$ cm$^{-3}$ ($-10\%$).

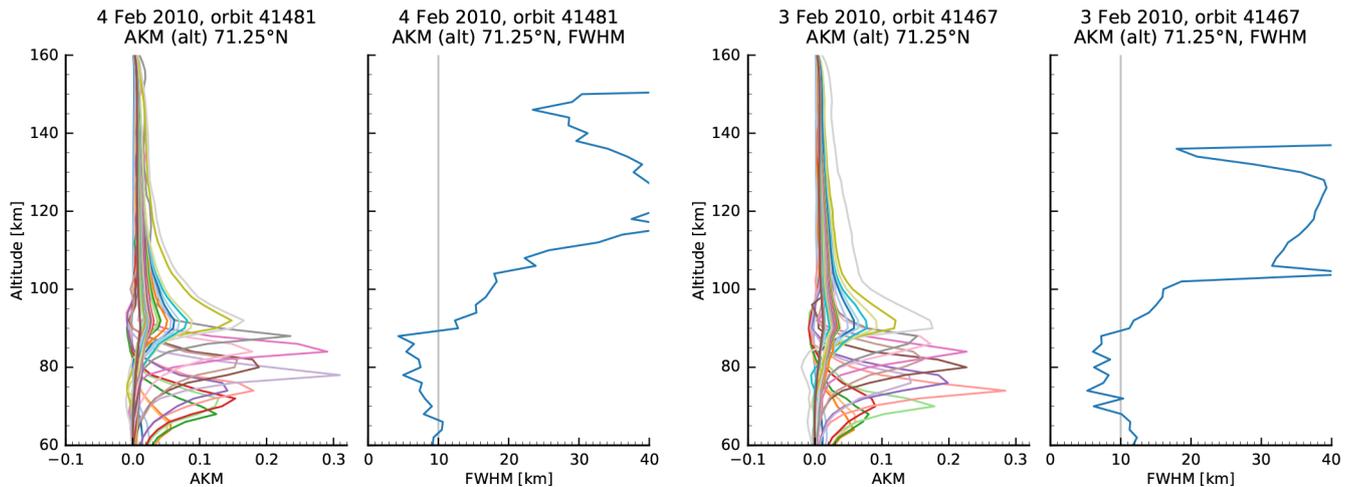

**Figure 9.** Altitude averaging kernel matrix elements (first panel from the left) and the respective full widths at half maximum (second panel) at the 71.25° N latitude bin for the sample nominal mode orbit (no. 41 481, 4 February 2010). For comparison, the averaging kernel matrix elements from the retrieval of the MLT scan limited to below 91 km and the full width at half maximum resolution at the same latitude bin of the sample MLT orbit are shown in the right two panels (sample orbit no. 41 467, 3 February 2010).

Since using the a priori values as they are seems to over-correct the missing data above 100 km, we extended the retrieval algorithm to include an additional scale factor $\alpha$ for the a priori values. This scale factor accounts for the highly variable NO column above 100 km. Instead of using $\alpha = 1$ and Eq. (3), we allow $\alpha$ to vary and use Eq. (5) to determine $x$ and $\alpha$ simultaneously.

The median of the number densities of the restricted MLT scans using the scaled a priori values compared to the number densities from the full MLT retrieval are shown in the left panel of Fig. 7. Using the fitted model number densities as a priori input brings the results much closer together between 60 and 90 km in all three regions. Some differences remain between 80 and 91 km at high southern latitudes, in particular when using the NOEM model as a priori input. The right panel of Fig. 7 shows the corresponding median of the relative differences.

At high northern latitudes, fitting the values of either model as a priori results in number density differences of less than 5 % below 85 km, and less than 10 % above. Fitting the NOEM model at middle and low latitudes results in differences of less than 3 % across the altitude range of interest. Using the fitted regression model results in a difference of less than 10 % up to around 80 km and up to 15 % larger number densities above. At high southern latitudes, the differences using the fitted NOEM model are the largest, up to $-25\%$ at the top, below $-20\%$ at 85 km and around $\pm 5\%$ below 75 km. Using the regression model here leads to differences below 10 % across the altitude range, below 85 km they are smaller than $\pm 6\%$, and below 75 km they are smaller than 2 %. Table summarises our findings regarding the different a priori choices.

## 4.4 Resolution

The vertical resolution using the SCIAMACHY MLT scans was extensively discussed in Bender et al. (2013). Here we briefly review the main results. The vertical averaging kernel elements of the MLT retrieval for the chosen example orbit are shown in the left panel of Fig. 8. As a measure of the vertical resolution, the corresponding full widths at half maximum (FWHM) are shown in the right panel of Fig. 8. From 70 to about 140 km, the vertical resolution is about 10 km or better.

Figure 9 (first panel from the left) shows the averaging kernel matrix elements from an example nominal mode orbit (no. 41 481 from 4 February 2010). The second panel in the same figure shows the vertical resolution given by the full width at half maximum of those line sums. The resolution is about 10 km or better between about 65 and 90 km and sharply decreases above 90 km. This is expected because the nominal mode limb scans deliver only column information below 91 km.

For comparison, the third and fourth panels of Fig. 9 show the averaging kernel matrix elements and their full widths at half maximum from restricting the MLT limb scans to 91 km. Between 60 and 90 km, the resolution resembles that of the full MLT scan shown in Fig. 8. The sharp drop in vertical resolution is directly related to the missing information above 91 km.

In terms of resolution, the above orbits are typical examples. For all retrievals the averaging kernels are similar in all regions where measurements are available, i.e. between ≈ 80° S and ≈ 75° N in northern winter and vice versa in northern summer.

## 5 Conclusions

We present an algorithm to retrieve NO number densities in the mesosphere from measured scattered solar radiation and atmospheric emissions in limb viewing geometry from the SCIAMACHY instrument on Envisat. We adapt and apply the method developed for SCIAMACHY's special MLT limb scans (50–150 km, Bender et al., 2013) to SCIAMACHY's nominal limb scans (−3 to 91 km).

The altitude range of plausible results from the SCIAMACHY nominal limb scans extends from 60 to about 85 km. The missing measurements above about 91 km manifest in larger NO number densities below 91 km if no prior input is used. Including a priori input in the retrieval leads to smaller number densities below 91 km for both our a priori choices, the NOEM model (Marsh et al., 2004) and the multilinear regression result from Bender et al. (2015). However, using an additional (free) scaling factor for the model inputs recovers the values below 91 km within 25 % and below 85 km within 20 % or better. Taking into account that the regression model already includes the SCIAMACHY MLT data, it should be used with caution. We therefore recommend the NOEM model values as prior input for the retrieval while simultaneously determining an appropriate scaling as the current best option. Unfortunately, comparisons between our NO data products and NO data products from other instruments are limited. The reason is that the sensitive altitude ranges are different – too low in some cases (e.g. Odin/SMR > 85 km) or too high in other cases (e.g. MIPAS nominal mode < 60 km). ACE-FTS, although measuring solar occultation, delivers NO data in the same altitude range. Comparisons are planned for the future after the full SCIAMACHY nominal mode data are processed and the ACE-FTS data have also been reprocessed.

The SCIAMACHY nominal limb scans provide almost 10 years of continuous daily spectra of the middle atmosphere (up to ≈ 90 km). Retrieving NO from these spectra is important for connecting solar activity to middle atmosphere composition and dynamics. Solar-wind and auroral particles precipitating into the upper atmosphere produce NO, which is reduced via photolysis by solar UV radiation. Therefore, NO has a long life in polar winter at auroral latitudes without direct sunlight. This allows NO to be transported downwards, sometimes to the stratosphere where it affects the polar ozone layer.

Ten years of daily measurements make it possible to investigate NO in the mesosphere with respect to long-term (years) and short-term (days) variations of solar activity (Sinnhuber et al., 2016). In particular, the production rate and lifetime of particle-produced NO in these regions can be estimated. Further work is planned to use this 10-year data set to constrain mesosphere NO in chemistry climate models as an important aspect of solar influence on climate.

## 6 Data availability

The SCIAMACHY/Envisat Level 1b spectra were downloaded via the official ESA data browser: https://earth.esa.int/web/guest/data-access/browse-data-products.

Downloading requires applying for data access at ESA. The calibration was carried out with the ESA tool `scial1c` available for download at https://earth.esa.int/web/guest/software-tools/content/-/article/scial1c-command-line-tool-4073. The retrieval code is available on request to the author (stefan.bender@kit.edu).

*Acknowledgement.* Stefan Bender and Miriam Sinnhuber thank the Helmholtz-society for funding this project under the grant number VH-NG-624. The SCIAMACHY project was a national contribution to the ESA Envisat, funded by German Aerospace (DLR), the Dutch Space Agency, SNO, and the Belgium ministry. The University of Bremen as Principal Investigator has led the scientific support and development of SCIAMACHY and the scientific exploitation of its data products. This study is also relevant to ESA studies such as MesosphEO. We acknowledge support by Deutsche Forschungsgemeinschaft and Open Access Publishing Fund of Karlsruhe Institute of Technology.

The article processing charges for this open-access publication were covered by a Research Centre of the Helmholtz Association.

Edited by: M. Rapp
Reviewed by: P. Espy and one anonymous referee